# Giant Magnetic Moments of Nitrogen Stabilized Mn Clusters and Their Relevance to Ferromagnetism in Mn Doped GaN

B. K. Rao and P. Jena
Physics Department, Virginia Commonwealth University
Richmond, VA 23284-2000

## Abstract

Using first principles calculations based on density functional theory, we show that the stability and magnetic properties of small Mn clusters can be fundamentally altered by the presence of nitrogen. Not only are their binding energies substantially enhanced, but also the coupling between the magnetic moments at Mn sites remains ferromagnetic irrespective of their size or shape. In addition, these nitrogen stabilized Mn clusters carry giant magnetic moments ranging from $4\mu_B$ in MnN to $22\mu_B$ in $Mn_5N$. It is suggested that the giant magnetic moments of $Mn_xN$ clusters may play a key role in the ferromagnetism of Mn doped GaN which exhibit a wide range (10K - 940K) of Curie temperatures.

Among all the elements in the 3d transition metal series, manganese is unique as an atom, cluster, crystal, or impurity. In this letter we show that the properties of small manganese clusters can be dramatically altered by introducing nitrogen. For example, otherwise weakly bound Mn clusters can bind strongly, their ferrimagnetic coupling could transform into ferromagnetic coupling resulting in giant magnetic moments, and the sensitivity of magnetic moments to cluster geometry can disappear. In addition, the nitrogen stabilized Mn clusters with their giant moments could play a major role in the observed ferromagnetism of Mn doped semi-conductors.

The atomic configuration of manganese is characterized by a half-filled 3d and filled 4s shell, namely $3d^5 4s^2$. The large energy gap (~8 eV) between $3d^5$ and $4s^2$ prevents significant s-d hybridization in clusters and crystals. Consequently the chemistry of manganese is influenced by the filled 4s shell and manganese atoms do not bind strongly with each other. As a matter of fact, $Mn_2$ is a weakly bonded Van der Waal's dimer with a binding energy[1] of 0.1 ± 0.1 eV to 0.56 ± 0.26 eV. Similarly, the cohesive energy of bulk Mn is the least among the 3d-transition metal elements. However, as an electron is removed from $Mn_2$, the resulting cation, namely $Mn_2^+$ binds strongly[2] and its inter-atomic distance of 3.06 Å is shorter than that of its neutral dimer, namely 3.4 Å. In contrast, the inter-atomic distances in bulk Mn vary between 2.25 Å and 2.95 Å. As Mn atoms begin to form clusters, their inter-atomic distances are around 2.9 Å and their binding energy per atom remains small, namely ~ 1.0 eV/atom.

The magnetic properties of Mn as clusters or crystals are also unique and are governed by their 3d electrons. The half-filled 3d shell, according to Hund's rule, yields an atomic magnetic moment of $5\mu_B$. The magnetic moments of $Mn_2$ are coupled antiferromagnetically[3] while $Mn_2^+$ is ferromagnetic with a total magnetic moment of $11\mu_B$. Small clusters of Mn containing up to 5 atoms and isolated in matrices exhibit ferromagnetic coupling with a magnetic moment of $5\mu_B$/atom[3]. As clusters become large, the coupling between Mn atoms become ferrimagnetic[4], and the large cancellation between majority and minority spins render net magnetic moments/atom which are significantly smaller than that in the free atom. This has been confirmed in recent experiments[5] where Mn clusters containing 11 to 100 atoms carry magnetic moments which are less than $1.5\mu_B$/atom.

As an impurity in semiconductors, Mn also exhibits unique properties. For example, recent discovery of ferromagnetism[6] in Mn doped InAs and GaAs and subsequent theoretical prediction[7] that the Curie temperature, $T_c$ of Mn doped GaN could be as high as the room temperature have created a lot of interest in the study of dilute magnetic semiconductors. In addition to the scientific interest in a fundamental understanding of the origin of ferromagnetism in these systems, the studies are driven by the potential technological merit of spin injection into wide-band gap materials.[8] The two main objectives of these studies are associated with increasing the Mn content and achieving a Curie point above the room temperature. Different experimental techniques have been tried and Curie temperatures ranging from 10 K to 940 K have been reported. For example, Overberg *et al*.[9] reported a Curie temperature between 10 and 25 K in GaN samples containing 7% Mn. Reed *et al*.[10] have been able to achieve a $T_c$ of 288-370 K by varying the growth and annealing conditions of Mn doped GaN. Recently Sonoda *et al*.[11] succeeded in incorporating up to 9% Mn in GaN and suggested a $T_c$ of 940 K. The reasons behind such a wide variation in the Curie temperatures are not known. Clearly the growth conditions are important as is the proper characterization of the sample. In this connection it is interesting to note that Overberg *et al*.[9], by using only the Mn cell and the nitrogen plasma under similar conditions, were able to grow samples that showed XRD peaks corresponding to the $Mn_4N$ stable phase. This phase is known to be ferrimagnetic[12] with a $T_c$ as high as 745 K.

Overberg et al.,[9] however, ruled out the possibility of the existence of $Mn_4N$ in the GaMnN film. We wonder if it is possible, under suitable growth conditions, to induce clustering of Mn around N in GaN, and if so, would such clusters carry large magnetic moments and cause high $T_c$'s?

The important question is whether Mn atoms can be made to bind more strongly, stabilize their ferromagnetic phase, and retain a large value for their individual magnetic moments - all at the same time. In this letter we show that this may indeed be possible. We find that Mn clusters can be substantially stabilized by nitrogen atoms by having their hybridized s-d electrons bond with the p electrons of nitrogen. This stabilization is accompanied by ferromagnetic coupling between the Mn atoms which, in turn, are anti-ferromagnetically coupled to N atoms. This nitrogen mediated ferromagnetic coupling also gives rise to giant magnetic moments of $Mn_xN$ clusters with total magnetic moments of $4\mu_B$, $9\mu_B$, $12\mu_B$, $17\mu_B$, and $22\mu_B$ for x = 1, 2, 3, 4, and 5 respectively. We also suggest that these giant "cluster magnets" may play a significant role in the observed ferromagnetism in Mn doped GaN semiconductors. In the following we give a brief outline of our theoretical procedure, a discussion of the results based on the bonding characteristics, and its relevance to our understanding of the origin of ferromagnetism in dilute magnetic semi-conductors.

The calculations are done using the molecular orbital theory where we represent the atomic orbitals of Mn and N atoms with the LANL2DZ gaussian basis available in the Gaussian 98 code.[13] The accuracy of this basis set has been tested[2] in small Mn clusters earlier against a more elaborate basis involving all electrons augmented with diffuse functions. While the equilibrium geometry and the inter-atomic distances do not differ, the binding energy/atom obtained using the LANL2DZ basis is overestimated compared to that using the all electron basis. However, the systematics, i.e. the energy gain in adding an atom to an existing cluster remains relatively unaffected. For example, the energy gains in going from $Mn_3$ to $Mn_4$ and from $Mn_4$ to $Mn_5$ are respectively 1.25 eV and 0.75 eV in the all-electron basis and 1.20 eV and 1.04 eV in the LANL2DZ basis. These results were based on the B3LYP form of the GGA. The total energies for a given cluster geometry are computed using the density functional theory and generalized gradient approximation (GGA) for exchange and correlation potential. We have used the BPW91 form for the GGA in the Gaussian code[13]. The geometries of the clusters are optimized without symmetry constraint by calculating the forces at every atom site and relaxing the geometry until the forces vanish. The threshold of these forces was set at 0.000102 a.u./Bohr. Several initial structures were tried to ensure that the globally optimized geometry does not correspond to a local minimum. For each cluster, we have also examined various spin multiplicities, $M = 2S+1$ to determine the magnetic moment of the ground state configuration.

In Fig. 1 we give the equilibrium geometries of $Mn_xN$ (x  5) clusters and compare these against pure $Mn_x$ clusters. No isomers lying close in energy to the ground state of $Mn_xN$ (x 3) were found. For $Mn_4N$ we found two isomers whose energies lay 0.10 eV and 0.30 eV above its ground state structure. For $Mn_5N$ we identified three isomers whose energies lay 0.54 eV, 0.72 eV, and 0.94 eV above the ground state. The geometries of these isomers along with their magnetic properties will be presented elsewhere. It is sufficient to mention here that the preferred spin multiplicities of the isomers were the same as their ground state structures. We note that the bond length of MnN dimer is rather small, namely 1.62Å. We have repeated the calculations of the MnN dimer using an all-electron (6-311G\*\*) basis and have obtained the optimized bond length of 1.59Å. As more Mn atoms are added, the structures assume two and three-dimensional shapes but the Mn-N distances change little from that in the dimer. The Mn-N-Mn bond angle in $Mn_2N$ is 96.1° and results from the interaction of the p orbitals of N with s-d hybridized orbitals of Mn. This nearly 90° bond angle between Mn-N-Mn remains a guiding rule for ground state geometries as we proceed to larger $Mn_xN$ clusters. For example, the angles between Mn-N-Mn in $Mn_3N$ cluster are all 90.7° and the structure is three-dimensional. The three Mn atoms form an equilateral triangle with the N atom 1.06Å above the plane. In $Mn_4N$ these angles are 89.1° as the N atom lays only 0.24Å above the $Mn_4$ square plane. The ground state structure of $Mn_5N$ is that of a triangular bipyramid of Mn atoms with one of the triangular faces capped by the N-atom. The Mn-Mn distances in all these clusters are about 2.8 ± 0.1 Å which are significantly shorter than the 3.4Å bond length of a $Mn_2$ dimer. We recall that in bulk Mn the inter-atomic distances vary between 2.25Å and 2.95Å. The Mn-N distances in Fig. 1 vary between 1.62Å and 1.96Å. We should point out that in the epitaxial film[14] of $Mn_3N_2$, the Mn-N distance is 2.1Å and the Mn-N-Mn angle is 90°.

The energetics of the clusters can be analyzed from the total energies of $Mn_xN$ and $Mn_x$ clusters corresponding to their respective ground state geometries. We have computed the atomization energies necessary to dissociate a cluster into individual atoms. $E_b^0$ and $E_b$ represent the atomization energies for $Mn_x$ and $Mn_xN$ clusters respectively. We have also calculated the energy gain, $\Delta$ in adding a Mn atom to an existing $Mn_{x-1}N$ cluster. Similarly $\Delta^0$ corresponds to the energy gain in adding a N atom to a $Mn_x$ cluster. These are defined as follows:

$$E_b^0 = [E(Mn_x) - xE(Mn)]/x \qquad (1)$$

$$E_b = - [E(Mn_xN) - xE(Mn) - E(N)]/(x+1) \qquad (2)$$

$$\Delta^0 = - [E(Mn_xN) - E(Mn_x) - E(N)] \qquad (3)$$

$$\Delta = - [E(Mn_xN) - E(Mn_{x-1}N) - E(Mn)] \qquad (4)$$

These results are given in Table 1.

We first discuss the energetics of the pure $Mn_x$ clusters, which also have been studied by previous authors.[2,15] We find that $Mn_2$ is not bound at the current level of theory. Previous calculations[2] using different levels of exchange-correlation functionals and basis sets have reported $Mn_2$ from being unbound to having a binding energy of as much as 1.5 eV. However, since $Mn_2$ is a weakly bound Van der Waals' dimer, as is evident from the low experimental binding energy, we can conclude that the results based on density functional theory cannot quantitatively account for the binding in $Mn_2$. The situation improves significantly in $Mn_3$ and larger clusters as the s and d electrons begin to hybridize. However, the binding is still weak as atomization energies of $Mn_3$, $Mn_4$, and $Mn_5$ are only 0.87 eV/atom, 1.16 eV/atom, and 1.22 eV/atom respectively.

As nitrogen atom is added, the binding energies of $Mn_xN$ clusters improve significantly. This can be seen by comparing the atomization energies of $Mn_xN$ clusters with those of $Mn_x$ clusters in Table 1. The energy gain in forming the MnN dimer using the LANL2DZ basis is 3.08 eV, which is in reasonable agreement with the all-electron (6-311G** basis) result of 2.59 eV. As pointed out earlier, the systematics in the energy again is not sensitive to the choice of basis sets and thus the following results based on LANL2DZ basis are considered reliable. Note that the atomization energy of $Mn_2N$ is 1.99 eV/atom while $Mn_2$ was not even bound. In larger clusters, the atomization energies of $Mn_xN$ clusters are significantly larger than those in pure $Mn_x$ clusters. A better way to demonstrate this is to examine the energy gained, $\Delta^0$ in adding a nitrogen atom to a $Mn_x$ cluster given in Table 1. Note that one gains as much as 6 eV in having a nitrogen atom to mediate the bonding among the Mn atoms. This can be understood clearly by examining bonding in $Mn_2N$ versus that in $Mn_2$. As noted earlier, the lack of hybridization between the s and the d electrons due to the filled 4s shell of Mn prevents $Mn_2$ from forming a strong bond. However, as a nitrogen atom is attached, the $4s^2$ electrons of manganese interact with the $2p^5$ electrons of nitrogen resulting in a strong bond. The charge transfer from Mn to N leaves the two Mn atoms in charged state and it is well known[2] that $Mn_2^+$ is a strongly bound dimer cation.

The next question is how many manganese atoms can be bound to a single nitrogen atom. For this it is instructive to analyze the value of $\Delta$, the energy gained in adding a manganese atom to a $Mn_{x-1}N$ cluster. These energies are also given in Table 1. Note that the energy gain, $\Delta$ decreases as the number of manganese atoms in the cluster increases. However, even for $Mn_5N$, it is still significant, namely 1.67 eV. Thus it is energetically favorable to certainly add at least five Mn atoms to a single nitrogen and this number could easily go further up. Other factors such as available phase space may also play an important role in determining the number Mn atoms that could cluster around a single nitrogen site in GaN. That clustering is energetically favorable does not seem to be in doubt.

Now we address the next important issue: Are these nitrogen stabilized Mn clusters magnetic and if so, what are their magnetic moments? As mentioned earlier, this was obtained by minimizing the total energy of each cluster with respect to all spin multiplicities (M = 2S+1). The total magnetic moments of these clusters corresponding to the ground state geometries are given in Table 1. The energetics of the neighboring spin states as well as the magnetic moment of the higher energy isomers will be published later. The magnetic moments of MnN and $Mn_2N$, where N is bonded to one or two Mn atoms, are respectively $4\mu_B$ and $9\mu_B$. These values can be represented by the formula, $(5x-1)\mu_B$ where x is the number of Mn atoms. The magnetic moments of $Mn_3N$, $Mn_4N$, and $Mn_5N$ are respectively $12\mu_B$, $17\mu_B$, and $22\mu_B$. These moments can be expressed by the formula, $(5x-3)\mu_B$. Note that in these structures N is bonded to three or four

Mn atoms. These large magnetic moments arise from ferromagnetic coupling between magnetic moments at Mn sites which are coupled antiferromagnetically with the magnetic moment at N site. The nature of this coupling is determined from the Mulliken population analysis of spins as well as by mapping the spin density distribution. The Mulliken analysis yields a spin magnetic moment of $4.30\mu_B$ at the Mn site and $-0.30\mu_B$ at the N site in MnN dimer. In $Mn_2N$, the magnetic moments at Mn and N sites are $4.54\mu_B$ and $-0.08\mu_B$ respectively. In the $Mn_3N$ cluster the Mulliken analysis yields a moment of $4.12\mu_B$ at each Mn site and $-0.36\mu_B$ at the N site. Similarly, the $17\mu_B$ moment of $Mn_4N$ results from $4.28\mu_B$ at each of the Mn sites and $-0.10\mu_B$ at the N site. The magnetic moments in $Mn_5N$ are $4.41\mu_B$ in the average at the Mn and $-0.07\mu_B$ at the N sites. In Fig. 2 we plot the spin density surfaces corresponding to a value of 0.005 a.u. Here the blue color represents positive and green represents negative spin densities. Note that the results in Fig. 2 are consistent with the above Mulliken population analysis.

From these cluster studies one can, therefore, conclude that nitrogen not only stabilizes the clustering of Mn atoms but also causes their magnetic moments to align ferromagnetically through a super-exchange type mechanism. The total magnetic moments of these clusters can be very large due to these interactions. What relevance can this have to the ferromagnetism of Mn-doped GaN and the variety of Curie temperatures that have been reported? We have carried out separate studies[16] of magnetic coupling between Mn atoms in GaN by studying clusters of $(GaN)_xMn_2$ as well as crystals of Mn-doped GaN. In the latter case two Mn atoms were substituted at different Ga sites using a 32 atom super cell. We have found the coupling to be ferromagnetic both in clusters and crystals with magnetic moments of $\sim3.5\mu_B$ at the Mn sites. Thus, Mn in GaN, whether forming clusters or substituted at Ga sites, tend to couple ferromagnetically.

We believe that the clustering of Mn around N could be responsible for the ferromagnetism of Mn doped GaN as well as the large variation in the Curie temperatures of different samples. This could arise from having nitrogen induced Mn clusters of different sizes in samples grown under different conditions. Note that the Curie Temperature of $Mn_4N$ is 745K. Thus, if Mn doped GaN, under suitable growth conditions, could contain $Mn_4N$ clusters, then their large magnetic moment, namely $17\mu_B$ could give rise to a large Curie temperature. On the other hand, if these clusters are small or Mn replaces a Ga atom, the small magnetic moment would then yield low Curie temperatures. This analysis suggests the importance of the growth mechanism. It is important to understand if clusters of Mn around N are, indeed, present in samples exhibiting large Curie temperature. EXAFS experiments could prove to be very useful in this regard. It may also be interesting to dope Mn into porous GaN which contains many defect sites and, thus, may precipitate Mn clustering.

This study also suggests that further investigations of clustering of Mn around As and oxygen may be useful. The former could illustrate if such clustering could lead to ferromagnetism of Mn-doped GaAs while the latter could suggest if Mn-doped ZnO may be a candidate, like Mn-doped GaN, for dilute magnetic semiconductors. Independent gas phase experiments involving Mn clustering in a N-seeded chamber can yield direct information on the magnetic character of $Mn_xN$ clusters. The ability to change magnetic coupling of Mn that carries a large atomic magnetic moment by chemical means would, indeed, be exciting. We hope that our study will stimulate such experiments.

We would like to thank Dr. G. P. Das for bringing to our attention the work of Overberg *et al.*[9] which prompted this investigation. We are grateful to Dr. A.K. Rajagopal for many stimulating discussions and a critical reading of the manuscript. This work was supported in part by a DURINT grant from the Office of Naval Research. The authors also acknowledge partial support by a grant (DEFG02-96ER45579) from the Department of Energy.

**Table 1**

Energetics of $Mn_xN$ clusters compared to those of $Mn_x$ clusters. Tabulated are the atomization energies, $E_b^0$, of $Mn_x$ clusters, atomization energies, $E_b$, of $Mn_xN$ clusters, energy gain, $\Delta^0$, in adding a nitrogen atom to a $Mn_x$ cluster, and energy gain, $\Delta$, in adding a manganese atom to $Mn_{x-1}N$ cluster. Also listed are the total magnetic moments of $Mn_xN$ clusters.

| x | $\mu_{total}$ ($\mu_B$) | $E_b^0$ (eV), Eq. (1) | $E_b$ (eV), Eq. (2) | $\Delta^0$ (eV), Eq. (3) | $\Delta$ (eV), Eq. (4) |
|---|---|---|---|---|---|
| 1 | 4 | - | 1.54 | 3.08 | 3.08 |
| 2 | 9 | - | 1.99 | 5.98 | 2.91 |
| 3 | 12 | 0.87 | 1.98 | 5.31 | 1.94 |
| 4 | 17 | 1.16 | 1.96 | 5.15 | 1.87 |
| 5 | 22 | 1.22 | 1.91 | 5.61 | 1.67 |

**Figure Captions**

1. Geometries of $Mn_x$ (left column) and $Mn_xN$ (right column) ($x \leq 5$) clusters in their ground states.

2. Spin density surfaces corresponding to 0.005 a.u. in $Mn_xN$ clusters. Figures a, b, c, d, and e correspond to x = 1, 2, 3, 4, and 5 respectively. The green surfaces represent negative spin densities around N site while the blue represents positive spin density around Mn sites.

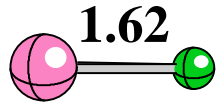
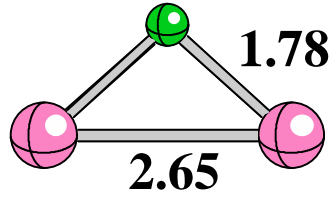
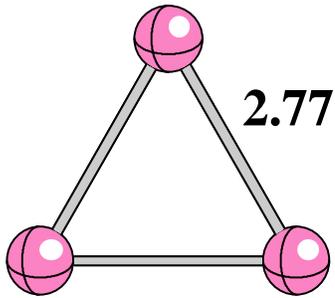
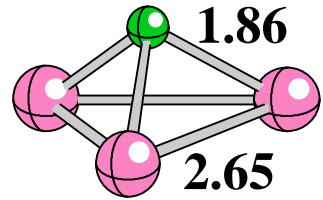
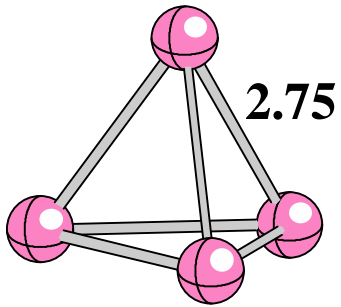
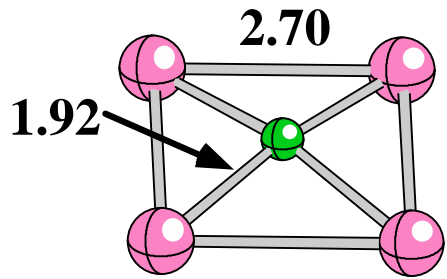
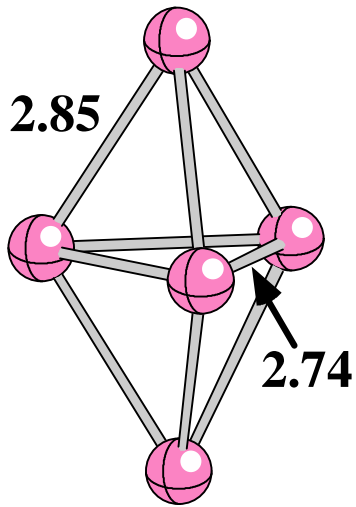
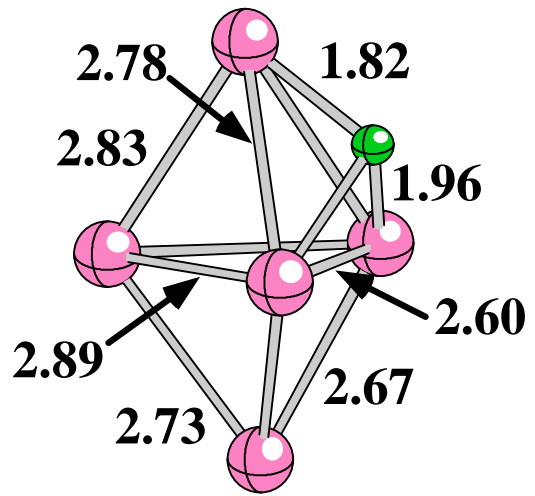

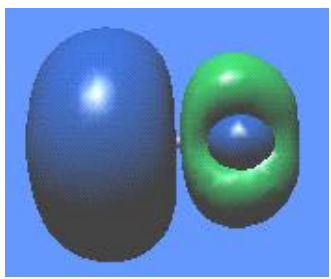

(a)

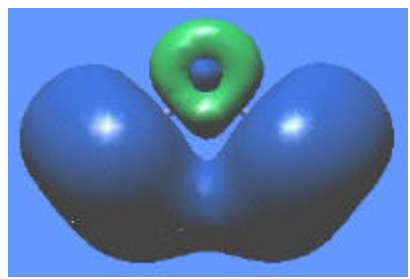

(b)

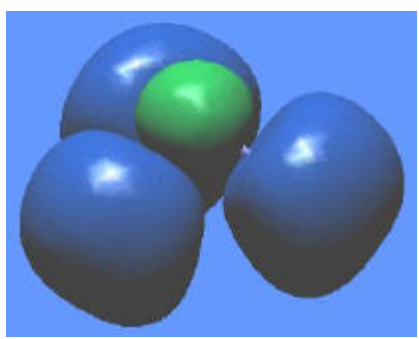

(c)

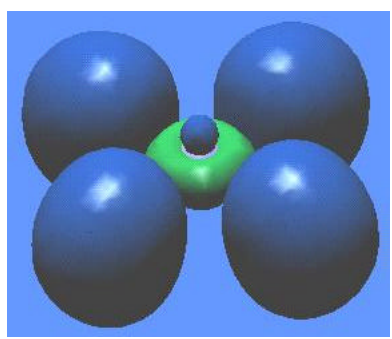

(d)

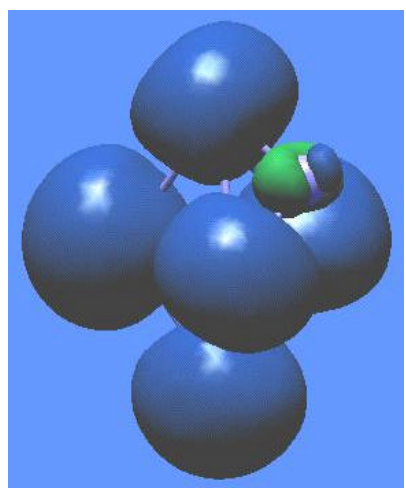

(e)